\documentclass[amsmath,superscriptaddress,reprint,aps]{revtex4-2}
\usepackage{graphicx}
\usepackage{dcolumn}
\usepackage{bm}
\usepackage{hyperref}
\usepackage{siunitx}
\usepackage{color, soul}

\usepackage{ulem}
\usepackage{xcolor}
\DeclareRobustCommand{\erase}{\bgroup\markoverwith{\textcolor{red}{\rule[.5ex]{2pt}{0.4pt}}}\ULon}

\setstcolor{red}

\begin{document}
	
	
	\title{Systematic dispersion engineering of crystalline microresonators for broadband and coherent frequency comb generation}

\author{Liu Yang}
\thanks{These authors contributed equally to this work}
\affiliation{Department of Physics, Faculty of Science and Technology, Keio University, 3-14-1 Hiyoshi, Kohoku-ku, Yokohama, 223-8522, Japan}
\affiliation{Department of Electronics and Electrical Engineering, Faculty of Science and Technology, Keio University, 3-14-1 Hiyoshi, Kohoku-ku, Yokohama, 223-8522, Japan}

 \author{Ryomei Takabayashi}
\thanks{These authors contributed equally to this work}
\affiliation{Department of Electronics and Electrical Engineering, Faculty of Science and Technology, Keio University, 3-14-1 Hiyoshi, Kohoku-ku, Yokohama, 223-8522, Japan}

 \author{Hiroki Moriguchi}
\thanks{These authors contributed equally to this work}
\affiliation{Department of Electronics and Electrical Engineering, Faculty of Science and Technology, Keio University, 3-14-1 Hiyoshi, Kohoku-ku, Yokohama, 223-8522, Japan}

\author{Hikaru Kodama}
\affiliation{Department of System Design Engineering, Faculty of Science and Technology, Keio University, 3-14-1 Hiyoshi, Kohoku-ku, Yokohama, 223-8522, Japan}

\author{Kazuma Miura}
\affiliation{Department of System Design Engineering, Faculty of Science and Technology, Keio University, 3-14-1 Hiyoshi, Kohoku-ku, Yokohama, 223-8522, Japan}

\author{Koshiro Wada}
\affiliation{Department of Electronics and Electrical Engineering, Faculty of Science and Technology, Keio University, 3-14-1 Hiyoshi, Kohoku-ku, Yokohama, 223-8522, Japan}

\author{Kai Yamaguchi}
\affiliation{Department of Electronics and Electrical Engineering, Faculty of Science and Technology, Keio University, 3-14-1 Hiyoshi, Kohoku-ku, Yokohama, 223-8522, Japan}

\author{Tatsuki Murakami}
\affiliation{Department of Physics, Faculty of Science and Technology, Keio University, 3-14-1 Hiyoshi, Kohoku-ku, Yokohama, 223-8522, Japan}

\author{Hajime Kumazaki}
\affiliation{Department of Physics, Faculty of Science and Technology, Keio University, 3-14-1 Hiyoshi, Kohoku-ku, Yokohama, 223-8522, Japan}
\affiliation{Department of Electronics and Electrical Engineering, Faculty of Science and Technology, Keio University, 3-14-1 Hiyoshi, Kohoku-ku, Yokohama, 223-8522, Japan}

\author{Yasuhiro Kakinuma}
\affiliation{Department of System Design Engineering, Faculty of Science and Technology, Keio University, 3-14-1 Hiyoshi, Kohoku-ku, Yokohama, 223-8522, Japan}

\author{Takasumi Tanabe}
\affiliation{Department of Electronics and Electrical Engineering, Faculty of Science and Technology, Keio University, 3-14-1 Hiyoshi, Kohoku-ku, Yokohama, 223-8522, Japan}

\author{Shun Fujii}
\email[]{shun.fujii@phys.keio.ac.jp}
\affiliation{Department of Physics, Faculty of Science and Technology, Keio University, 3-14-1 Hiyoshi, Kohoku-ku, Yokohama, 223-8522, Japan}

	
\begin{abstract}
Ultraprecision machining offers a powerful route to dispersion control in crystalline microresonators, allowing the design of waveguide geometries for tailoring the spectrum of microresonator frequency combs. By precisely designing the geometry, both group-velocity and higher-order dispersions can be engineered across a broad wavelength range. However, despite their promising features, such advantages have remained largely unexplored due to fabrication challenges. Here, we demonstrate that resonators shaped by ultrapecision machining exhibit high precision and strongly suppressed spatial mode interactions, facilitating the generation of smooth dissipative Kerr soliton combs and broadband frequency combs beyond the telecommunication C-band. These results underscore the effectiveness of precision geometry control for realizing coherent and broadband microcombs on crystalline photonic platforms.
\end{abstract}

\maketitle
\section{Introduction}
Optical microresonators with high quality (Q) factors and small mode volumes significantly reduce the threshold power required for parametric oscillations, making them highly suitable for optical frequency comb generation~\cite{Kippenberg2018,doi:10.1126/science.aay3676}. In particular, ultrahigh-Q whispering gallery mode (WGM) crystalline microresonators, which exhibit ultralow losses and high finesse~\cite{GRUDININ200633,Savchenkov2007}, enable the realization of narrow-linewidth lasers~\cite{Liang2015a} and the observation of nonlinear optical phenomena at extremely low power levels~\cite{Kippenberg2004}. Since the first demonstration of dissipative Kerr soliton (DKS) microcombs using crystalline microresonators~\cite{Herr2014}, DKSs have exhibited excellent features for practical applications, including spectroscopy~\cite{Suh2016}, optical clocks~\cite{Wu2025}, astronomical calibration~\cite{Obrzud2019}, high-capacity communications~\cite{Fueloep2018}, low-noise frequency synthesis~\cite{Spencer2018}, ranging~\cite{Trocha2018}, and photonic convolutional processing~\cite{Feldmann2021}. Despite exceptional advances in photonic integrated platforms in the recent years~\cite{Gaeta2019}, crystalline microresonators still offer narrower cavity linewidths and less thermo-refractive noise than other microresonator platforms~\cite{Lim2017}. These features are advantageous for low-phase-noise DKS generation~\cite{Liang2015,Lucas2020,Murakami2025} and the implementation of compact reference cavities~\cite{Alnis2011,Jin2025}, and are also highly compatible with self-injection locking techniques~\cite{Kondratiev2023}.

The dispersion of microresonators primarily governs the dynamics of frequency comb generation, as anomalous group-velocity dispersion is essential for triggering modulation instability, namely, parametric gain under continuous-wave (CW) pumping~\cite{Leo2010}. A key distinction between microresonator optics and fiber optics is the complexity of the resonance modes, which can interact with one another.  Moreover, such mode interactions often appear as local perturbations in multimode waveguides, leading to unique DKS phenomena such as soliton breathing~\cite{Lucas2017a}, soliton molecules~\cite{Weng2020}, soliton crystals~\cite{Cole2017}, and dispersive wave emission~\cite{Yi2017}. Strong mode interactions result in avoided mode crossings (AMXs) that have also been exploited to achieve dark pulses~\cite{Xue2015}, highly efficient microcomb generation~\cite{Helgason2023}, and, in some cases, microcomb stabilization~\cite{Bao2017}. However, excessive AMXs can inhibit soliton formation~\cite{Herr2014b} or cause random spectral distortions in the soliton envelope~\cite{Yi2015,Matsko2016a}. These trends are more prominent in manually fabricated crystalline WGM resonators, where uncontrollable AMXs occur throughout the spectra~\cite{Liu2022,Qu2024}.

Precise control over resonator geometry allows us to tailor its dispersion characteristics, a process known as dispersion engineering~\cite{FujiiTanabe+2020+1087+1104}. In recent years, dispersion engineering has been widely applied to CMOS-compatible microresonator platforms due to their high design flexibility and mature fabrication processes~\cite{Kim2017,Li2023,Ye2023}. However, dispersion engineering is more challenging in crystalline resonators, where the available geometrical degrees of freedom are often limited by fabrication constraints.  In particular, ultrahigh-Q crystalline microresonators are typically fabricated by mechanical shaping and polishing~\cite{Qu2023,Fujii2023,Fujii2024}. This process limits geometric precision, thereby hindering reproducibility and accurate dispersion engineering. Moreover, it often produces microresonators supporting a large number of spatial modes, which greatly complicate the nonlinear dynamics through mode coupling~\cite{Matsko2016a,Yang2016b}. To address these issues, computer-controlled ultraprecision machining techniques have been widely employed~\cite{Grudinin2015,Nakagawa2016,Pavlov2017,Grudinin2017,Fujii:20}. This enables the fabrication of microresonators with pre-designed mode structures, facilitating the reproducibility and dispersion engineering that are crucial for microresonator frequency comb generation. Despite these potential advantages of ultraprecision machining, neither  microresonator dispersion controllability nor design guidelines for crystalline WGM waveguides for DKS generation have been fully established.

In this paper, we present a systematic approach to the dispersion engineering of ultrahigh-Q crystalline microresonators for broadband, coherent Kerr frequency comb generation. We first establish general design guidelines for achieving anomalous dispersion in magnesium fluoride ($\mathrm{MgF_2}$) resonators based on both geometric parameters and material properties. We then experimentally demonstrate that ultraprecision machining significantly suppresses spatial mode interactions and enables mode-interaction-free soliton microcomb generation. Furthermore, we show that various waveguide geometries can be implemented via computer-controlled machining to finely tailor the dispersion profile. This allows for flexible dispersion control in the O and E telecommunication bands as well as in the mid-infrared region, and even enables widely tunable optical parametric oscillation (OPO) at a pump wavelength of 1~\textmu m  through higher-order dispersion engineering. These results provide a comprehensive strategy for the design and fabrication of crystalline microresonators toward versatile and application-oriented microcomb platforms.

\begin{figure*}
\includegraphics[width=1\linewidth]{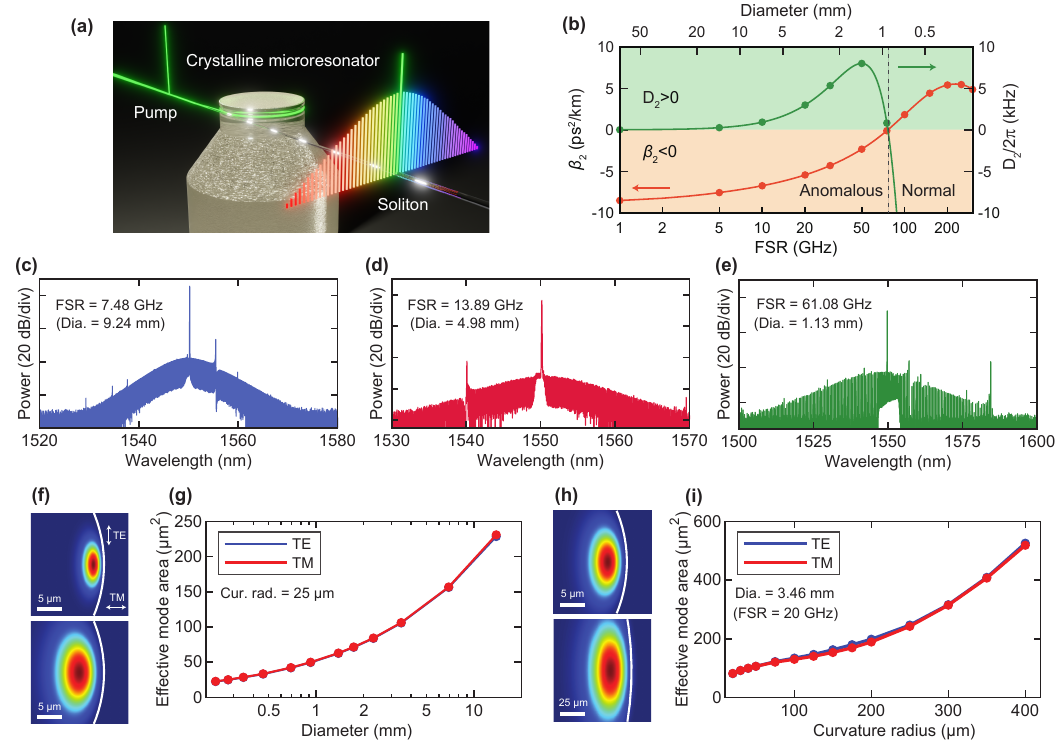}
\caption{\label{Fig_schematics} (a) Schematic illustration of a soliton microcomb in a crystalline WGM microresonator. (b) Group velocity dispersion $\beta_2$ and $D_2/2\pi$ values of $\mathrm{MgF_2}$ resonators for different FSRs. All simulations were performed for a fundamental mode. (c-e) Experimentally observed single soliton spectra with FSRs of 7.48~GHz, 13.89~GHz, and 61.08~GHz, respectively. (f) Mode field distributions with FSRs of 100~GHz (top) and 10~GHz (bottom) under a fixed curvature radius of 25~\textmu m. (g) Effective mode area as a function of the resonator diameter for TE and TM modes with a curvature radius of 25~\textmu m. (h) Mode field distributions with curvature radii of 25~\textmu m (top) and 100~\textmu m  (bottom) under a fixed FSR of 25~GHz. (i) Effective mode area as a function of the curvature radius for TE and TM modes with an FSR of 20~GHz.}
\end{figure*}

\begin{figure*}
\includegraphics[width=0.9\linewidth]{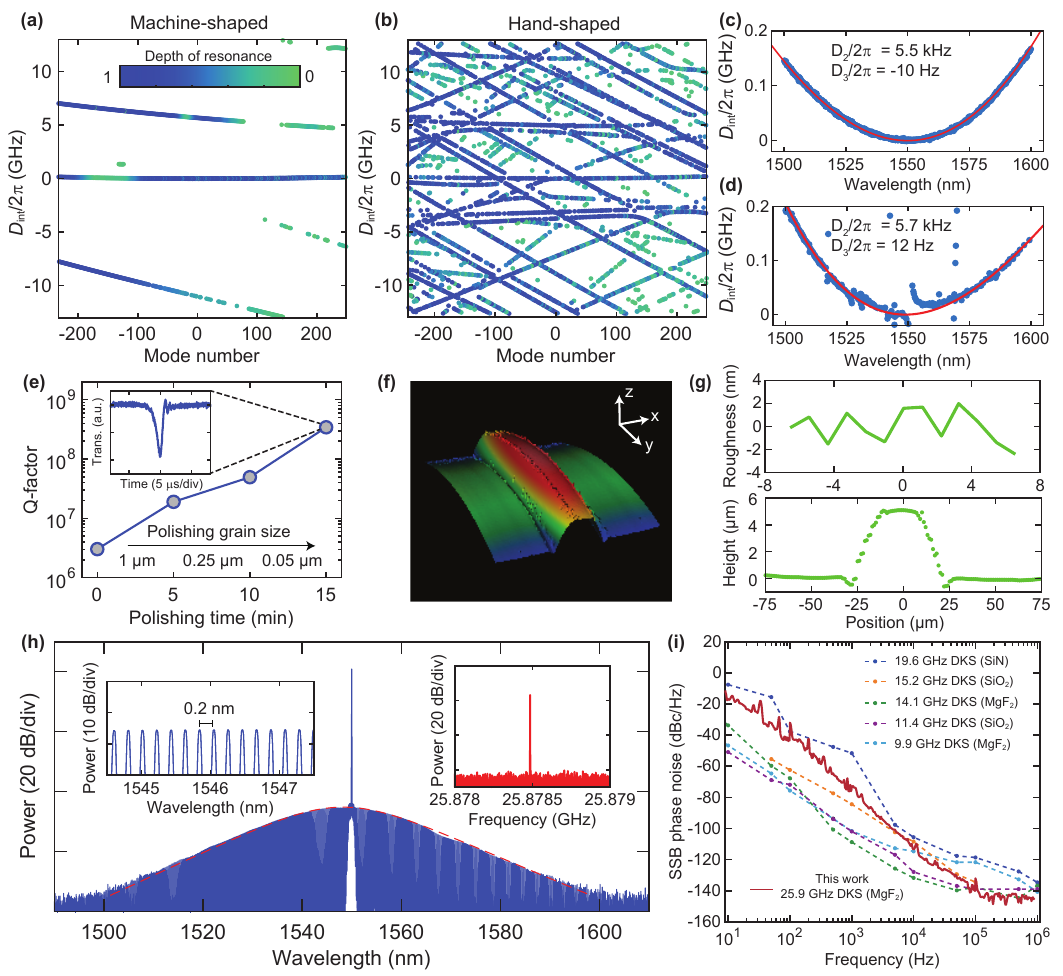}
\caption{\label{Fig_soliton} (a,b) Mode structures of 25~GHz-FSR $\mathrm{MgF_2}$ microresonators fabricated by machining and hand-shaping. A significant reduction in the number of mode families is confirmed for a machine-shaped resonator. (c,d) Magnified plots (blue) and fitting curves (red) for the machine-shaped (c) and the hand-shaped (d) resonators. A large deviation from a parabolic curve due to mode coupling is observed only in a hand-shaped resonator. (e) The Q-factors gradually improved with additional polishing. (f) 3D profile of the machine-shaped microresonator. (g) Measured surface roughness (top) and the cross-sectional shape (bottom) of the resonator. (h) A single soliton spectrum with a smooth $\mathrm{sech^2}$ envelope generated in a machine-shaped resonator. The fitting result yields a 3-dB bandwidth of 2.61~THz, corresponding to a pulsewidth of 121~fs. The insets show the magnified optical spectrum and the beat-note spectrum, yielding a repetition rate of 25.88~GHz. (i) Phase noise comparison with other free-running DKS-based photonic microwaves scaled to 25.9~GHz. Performances of a 19.6~GHz DKS in $\mathrm{Si_3N_4}$~\cite{Liu2020NP}, a 15.2~GHz DKS in $\mathrm{SiO_2}$~\cite{Yang2021}, a 14.1~GHz DKS in $\mathrm{MgF_2}$~\cite{PhysRevLett.122.013902}, an 11.4~GHz DKS in $\mathrm{SiO_2}$~\cite{Yao2022}, a 9.9~GHz DKS in $\mathrm{MgF_2}$~\cite{Liang2015}, and a 25.9~GHz DKS in $\mathrm{MgF_2}$ (this work).}
\end{figure*}

\section{General guidelines for dispersion and mode engineering in magnesium fluoride microresonators}

Microresonator dispersion is one of the dominant factors governing dissipative Kerr solitons, as it must counterbalance Kerr nonlinearities (self- and cross-phase modulation) to sustain ultrashort pulse waveforms in optical microresonators. The overall dispersion arises from both material and geometric contributions, and anomalous dispersion, namely negative group velocity dispersion (GVD) $\beta_2 < 0$ is needed to obtain a soliton spectrum whose envelope closely follows a $\sim\mathrm{sech}^2$ profile~\cite{Herr2014}. Figure~\ref{Fig_schematics}(a) shows a schematic illustration of a magnesium fluoride ($\mathrm{MgF_2}$) crystalline microresonator and the optical spectrum of soliton microcombs pumped by a CW laser.

Microresonator dispersion can be accurately simulated using a finite-element method (FEM), which provides an eigenvalue solver for axisymmetric structures~\cite{Oxborrow2007}. In this work, unless otherwise noted, we use commercially available software (COMSOL Multiphysics) to simulate mode profiles and microresonator dispersion. While several geometric parameters contribute to the overall dispersion, the resonator diameter has a major impact on the GVD in WGM microresonators. Figure~\ref{Fig_schematics}(b) shows the effect of the resonator FSR, which is inversely proportional to the resonator diameter, on the dispersion $\beta_2$. It is convenient to express microresonator dispersion using coefficients derived from a Taylor expansion of the resonant frequencies $\omega_\mu$~\cite{Herr2014b,FujiiTanabe+2020+1087+1104}, yielding the second-order dispersion as $D_2 = -cD_1^2 \beta_2/n$, where $D_1/2\pi$ corresponds to the equidistant FSR in the pump mode ($\mu=0$), and $n$ denotes the refractive index of the material. The result shows that an FSR of 75~GHz or lower, corresponding to a diameter of approximately 900~\textmu m or larger, leads to anomalous dispersion. This boundary generally determines the viable size range of $\mathrm{MgF_2}$ resonators suitable for soliton generation.  Figures~\ref{Fig_schematics}(c)-\ref{Fig_schematics}(e) show the optical spectra of single soliton states with three different FSRs. As these resonators naturally exhibit moderate anomalous dispersion, we readily find mode families that support soliton states. It should be noted that an auxiliary laser method~\cite{Zhang2019a,Niu2024} is employed to mitigate strong thermal effects in the smallest resonator, namely one with an FSR of 61~GHz.

When the WGM resonator is smaller or the pump wavelength deviates significantly from the telecom C-band, certain tricks are typically required to enable soliton comb generation. One promising approach is to engineer the waveguide structure rather than simply to adjust the resonator diameter. In weakly confined cases, such as spheroidal geometries where the WGM is loosely confined by the waveguide, the curvature radius has limited influence on the dispersion. However, the effect of the waveguide structure becomes more pronounced when the mode is tightly confined, and the optical field is strongly perturbed by the surrounding geometry. In general, rectangular structures tend to enhance anomalous dispersion, whereas triangular structures promote normal dispersion. These trends can be attributed to the changes in both the effective radius and the effective mode index as functions of wavelength that are unique to WGM resonators~\cite{FujiiTanabe+2020+1087+1104}.

The resonator geometry also influences the effective mode area of the resonator. Figures~\ref{Fig_schematics}(f)-\ref{Fig_schematics}(i) show the effective mode area as a function of diameter and curvature radius, along with the corresponding mode distributions. Since the power threshold of a four-wave mixing (FWM) process is inversely proportional to the mode volume, reducing the curvature radius reduces the pump power required for soliton comb generation. A larger mode area, however, can suppress thermo-refractive noise, which is a fundamental noise source in soliton combs, resulting in a trade-off between threshold power and soliton phase noise~\cite{Yao2022}.

Because uniaxial $\mathrm{MgF_2}$ crystals exhibit birefringence, transverse-electric (TE) and transverse-magnetic (TM) modes have slightly different refractive indices ($\sim$1.38 for TE and $\sim$1.37 for TM) in the $z$-cut configuration. According to waveguide theory, TM modes possess a strong electric field component oriented in the azimuthal direction (i.e., the propagation direction), whereas TE modes exhibit a very weak component in that direction. In the cross-sectional view of the mode field (Fig.~\ref{Fig_schematics}(f) and \ref{Fig_schematics}(h)), the electric field is oriented vertically for TE modes and horizontally for TM modes in the spheroidal WGM structure. Importantly, this relation reverses when the waveguide becomes a thin disk structure similar to a slab waveguide, wherein the azimuthal component is maximized when the electric field is oriented vertically, and thus, we define this as the TM mode. Although mode polarization has little effect on the effective mode area, it plays a critical role in mode hybridization~\cite{Ramelow2014,Lee2017} and higher-order dispersion~\cite{Yang2016}.

\section{Mode engineering by ultraprecision machining for mode-interaction-free soliton combs}

Crystalline WGM resonators are generally fabricated through shaping and polishing. Single-point diamond turning using a machining center is a reliable shaping technique that enables the fabrication of multi-resonator stacks~\cite{Pavlov2018} and microstructures~\cite{Grudinin2015,Nakagawa2016,Fujii:20}. Although ultraprecision machining alone can achieve Q-factors of up to 100 million by optimizing the cutting conditions~\cite{Fujii:20,Hayama2022}, additional polishing is typically required to reach ultrahigh-Q levels. Polishing with diamond slurry or sandpaper significantly reduces surface roughness to the sub-nanometer level, enabling Q-values of the order of $10^8-10^{10}$~\cite{Savchenkov2007}.

Here, we highlight a potential advantage of ultraprecision machining in suppressing spatial mode interactions and the resulting avoided mode crossings (AMXs). Mode interactions between different mode families are inevitable in overmoded resonators, and their effects are manifested as spectral peaks and dips, as observed in Figs.~\ref{Fig_schematics}(c)-\ref{Fig_schematics}(e).
 
Figures~\ref{Fig_soliton}(a)–\ref{Fig_soliton}(d) compare the mode structures and magnified dispersion profiles of machine-shaped (Figs.~\ref{Fig_soliton}(a) and \ref{Fig_soliton}(c)) and hand-shaped (Figs.~\ref{Fig_soliton}(b) and \ref{Fig_soliton}(d)) resonators, each with an FSR of $\sim$25~GHz. A significant reduction in the number of mode families and spatial mode interactions is clearly observed. This is primarily because ultraprecision machining produces a small curvature radius, which leads to a reduction in the resonator's cross-sectional area. As a result, only a few mode families are observed in the transmission spectrum. The relatively low Q-factors of the initial machine-shaped resonators ($\sim$$10^7$) can be readily improved through mechanical polishing. The resonance mode analyzed here belongs to the same mode family as that shown in Fig.~\ref{Fig_soliton}(c) and is the identical resonance mode used for soliton generation described later. This process is summarized in Fig.~\ref{Fig_soliton}(e). We also measure the resonator shape using a scanning white-light interferometer (NewView 6200, Zygo) as shown in Fig.~\ref{Fig_soliton}(f). The retrieved cross-sectional shape and the surface roughness along $x$-axis are presented in Fig.~\ref{Fig_soliton}(g), yielding only $\sim$5~\textmu m waveguide height and the average surface roughness (Sa) of 1.8~nm in the measurement area of 25~\textmu m$\times$15~\textmu m. Such a tightly confined and smooth microstructure leads to a substantial reduction in both the number of mode families and the mode area, which are beneficial for soliton microcomb generation. It should also be noted that surface roughness at the $\sim$2~nm level corresponds to scattering-limited Q-factors on the order of $10^8-10^9$~\cite{Fujii:20}, consistent with the Q-values observed in our experiments.  

After improving the Q-factors, we excite the selected mode with a pump power of several tens of milliwatts. A DKS microcomb is initiated by rapidly scanning the pump laser frequency across the resonance~\cite{Herr2014,Fujii2023}. The observed single-soliton spectrum exhibits no noticeable spectral distortions caused by AMXs, consistent with the ideal dispersion profile (Fig.~\ref{Fig_soliton}(h)). The spectrum has a 3-dB bandwidth of 2.61~THz, corresponding to a pulse width of 121~fs and a repetition rate of 25.88~GHz. These results indicate that ultraprecision machining is a powerful tool for obtaining smooth, mode-interaction-free soliton envelopes with low pump powers.

To confirm the low-noise performance of the observed DKS comb, we measure a single-sideband (SSB) phase noise and compare the performance with other free-running DKS-based photonic microwaves at the X-, Ku- and K-bands (all scaled to 25.9~GHz). As shown in Fig.~\ref{Fig_soliton}(i), our result shows the lowest phase noise, particularly at high offset frequencies above 100~kHz, which indicates significant instantaneous phase stability. We attribute the phase noise degradation at low offset frequencies to temperature fluctuations, environmental perturbations, and detuning instabilities. The stability in this region can be disciplined by robust packaging~\cite{Yang2021,Yao2022}, microwave injection locking~\cite{PhysRevLett.122.013902,Liu2020NP} and quiet-point operation~\cite{Yang2021,Yao2022,Lucas2020}.

\section{Systematic dispersion engineering by ultraprecision machining}

\begin{figure*}
\includegraphics[width=0.9\linewidth]{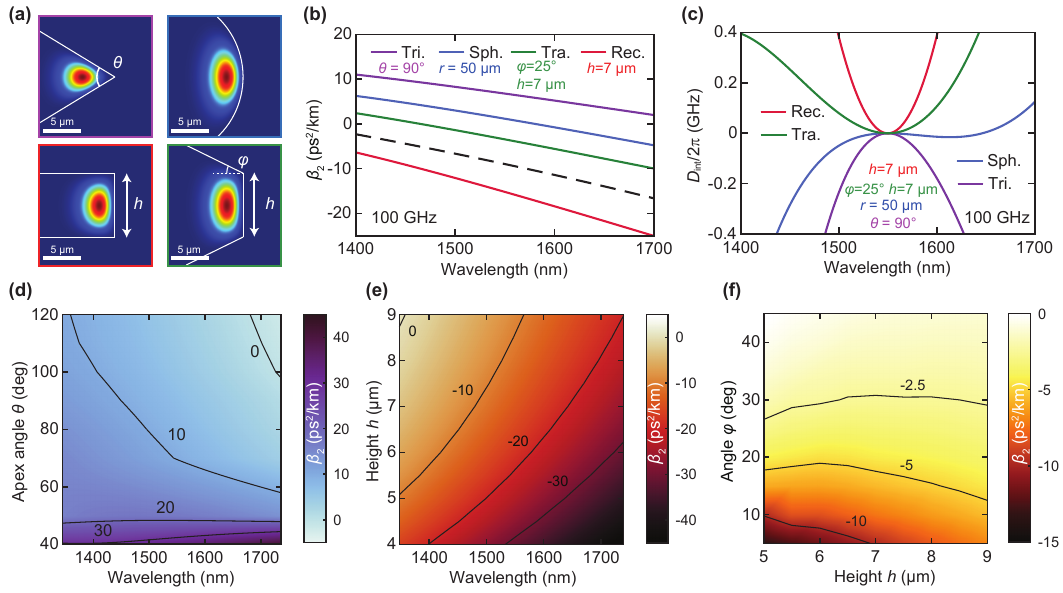}
\caption{\label{Fig_dispersion_1550} (a) Mode field distributions for different WG structures of interest. The scale bar represents 5~\textmu m. (b,c) Simulated group velocity dispersion $\beta_2$ and corresponding integrated dispersion $D_\mathrm{int}/2\pi$ for 100~GHz FSR resonators. The dashed black line is the material dispersion of an $\mathrm{MgF_2}$ crystal for reference. (d,e) Contour maps for GVDs showing the dependence on the apex angle $\theta$ for triangular shapes and the height $h$ for rectangular shapes. (f) Contour map for GVDs as a function of the angle $\varphi$ and the height $h$ of trapezoidal shapes. The 100 GHz FSR is assumed for (d-f).}
\end{figure*}

Another advantage of computer-controlled machining is its ability to fabricate sophisticated waveguide structures as designed. A spheroid-shaped $\mathrm{MgF_2}$ resonator with an FSR greater than 75~GHz exhibits normal dispersion, which prevents bright soliton generation. Nevertheless, there remains a demand for soliton combs with such large FSRs in specific applications such as optical communications~\cite{Corcoran2020,Fujii:22} and low-noise W-band millimeter-wave generation~\cite{Savchenkov2024}. To meet this requirement, dispersion engineering via precision machining is both challenging and essential.

We consider four different resonator geometries, as illustrated in Fig.~\ref{Fig_dispersion_1550}(a). Figures~\ref{Fig_dispersion_1550}(b) and \ref{Fig_dispersion_1550}(c) show simulated group-velocity dispersion (GVD) $\beta_2$ and integrated dispersion $D_\mathrm{int} (= D_2\mu^2/2 + D_3\mu^3/6 + \cdots)$ for $\mathrm{MgF_2}$ resonators with an FSR of 100~GHz. The spheroidal shape exhibits normal dispersion, whereas the other structures show distinct and unique dispersion trends. The results indicate that the triangular shape enhances normal dispersion. The dependence on the apex angle is shown in Fig.~\ref{Fig_dispersion_1550}(d), where sharper apex angles are found to strongly promote normal dispersion. This behavior can be explained by the rapid decrease in the effective mode radius with wavelength, a phenomenon also observed with wedged disk resonators~\cite{Lee2012}. The rectangular shape exhibits a dispersion trend distinct from that of the triangular geometry. As the waveguide height decreases, the anomalous dispersion becomes stronger due to the tight mode confinement effect, as shown in Fig.~\ref{Fig_dispersion_1550}(e). The trapezoidal shape offers a compromise between these features, and its dispersion can be flexibly tuned by adjusting two structural parameters, namely the angle $\varphi$ and the height $h$. Figure~\ref{Fig_dispersion_1550}(f) shows the dispersion at 1550~nm for these parameters, demonstrating that a suitable anomalous dispersion range can be achieved for soliton generation.

\begin{figure*}
\includegraphics[width=0.9\linewidth]{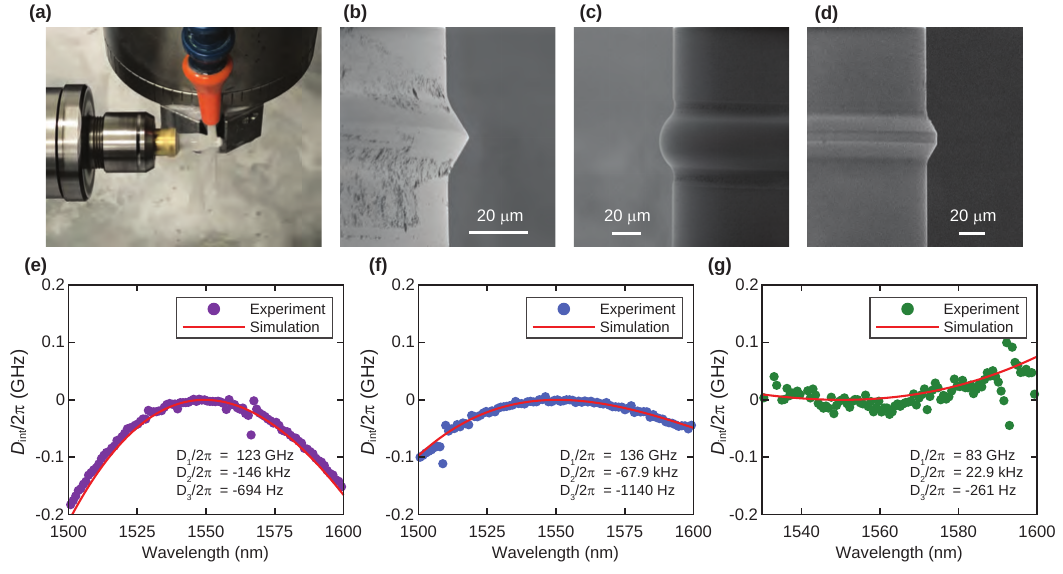}
\caption{\label{Fig_dispersion} (a) Ultraprecision lathe for microresonator fabrication. (b-d) SEM images for $\mathrm{MgF_2}$ resonators fabricated by fully computer-controlled machining. (e-g) Measured and simulated dispersion $D_\mathrm{int}/2\pi$. A triangular shape with an apex angle of 120$^\circ$ for (b,e). A spheroid shape with a curvature radius of 36~\textmu m for (c,f). A trapezoidal shape with $h=$8~\textmu m and $\varphi=45^\circ$ for (d,g). }
\end{figure*}

Ultraprecision machining enables the formation of microstructured waveguides with high reproducibility. Here, we fabricate three resonators with different waveguide structures using an ultraprecision aspheric surface machine (ULG-100E, Toshiba Machine Ltd.), as shown in Fig.~\ref{Fig_dispersion}(a). The fabrication procedure is detailed in Ref.~\cite{Fujii:20}. Figures~\ref{Fig_dispersion}(b)-\ref{Fig_dispersion}(d) show scanning electron microscope (SEM) images of the fabricated resonators. The measured and simulated dispersions are plotted in Figs.~\ref{Fig_dispersion}(e)-\ref{Fig_dispersion}(g). The design parameters are as follows: a triangular shape with an apex angle of $\theta = 120^\circ$ (Figs.~\ref{Fig_dispersion}(b) and \ref{Fig_dispersion}(e)); a spheroidal shape with a curvature radius of $r = 36$~\textmu m (Figs.~\ref{Fig_dispersion}(c) and \ref{Fig_dispersion}(f)); and a trapezoidal shape with $h = 8$~\textmu m and $\varphi = 45^\circ$ (Figs.~\ref{Fig_dispersion}(d) and \ref{Fig_dispersion}(g)). The measured dispersion agrees well with the simulated results, demonstrating that ultraprecision turning allows us to obtain both the resonator shape and its dispersion as designed. The dispersion measurements are performed through precise frequency calibration with a fiber Mach–Zehnder interferometer~\cite{FujiiTanabe+2020+1087+1104}. Although typical Q-factors are limited to $10^6$-$10^7$ at the current stage due to possible additional scattering losses at the edges, these values can be improved by further optimizing the machining conditions or by employing additional polishing.

\section{Dispersion engineering strategy for microcomb generation in O-, E-telecom bands, and mid-infrared region}

\begin{figure*}
\includegraphics[width=0.9\linewidth]{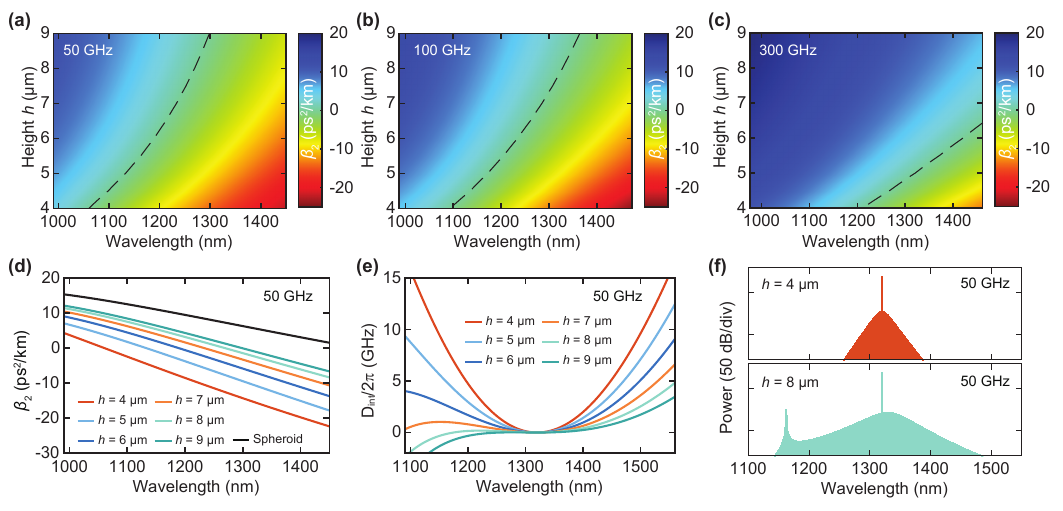}
\caption{\label{Fig_dispersion_1300} (a-c) Colormaps of GVDs with respect to rectangular height $h$ for 50, 100, and 300~GHz-FSRs at $\sim$1300~nm. The dashed black lines represent zero dispersion boundaries. (d,e) Simulated GVD $\beta_2$ and corresponding integrated dispersion $D_\mathrm{int}/2\pi$ for 50~GHz FSR resonators. The black line indicates the GVD of the spheroid shape with a curvature $r$=25~\textmu m for comparison. (f) Simulated comb spectra for $h=4$~\textmu m and $h=8$~\textmu m. The LLE is numerically solved to produce optical spectra including higher-order dispersions.}
\end{figure*}

Beyond the SCL-band (1460–1625~nm), shorter-wavelength telecom bands such as the O-band (1260–1360~nm) and E-band (1360–1460~nm) introduce strong normal dispersion in microresonators. As shown in Fig.~\ref{Fig_dispersion_1550}, rectangular geometries enable the compensation of material dispersion, and the same strategy can be applied here. Figures~\ref{Fig_dispersion_1300}(a)-\ref{Fig_dispersion_1300}(c) show contour maps of the GVD as a function of the waveguide height $h$ for various FSRs. These results reveal a gradual transition from normal to anomalous dispersion as the waveguide height $h$ decreases. Figures~\ref{Fig_dispersion_1300}(d) and \ref{Fig_dispersion_1300}(e) highlight the case for a 50~GHz FSR. By varying only the waveguide height, the GVD can be tuned from near-zero to strongly anomalous values. This tunability directly affects the bandwidth of the soliton microcombs, as shown in Fig.~\ref{Fig_dispersion_1300}(f). Simulated soliton spectra based on the mean-field Lugiato–Lefever equation (LLE)~\cite{Coen2013} show that a resonant dispersive wave~\cite{Peng2008} appears at 1180~nm, extending the optical spectrum for $h = 9$~\textmu m, whereas a resonator with $h=4$~\textmu m yields a narrow-bandwidth soliton spectrum. The LLE simulation is conducted using the Split-step Fourier method including higher-order dispersion~\cite{Brasch2016a}. The numerical simulation is detailed in Appendix~A.

In contrast to shorter wavelengths, where the material dispersion exhibits normal GVD, $\mathrm{MgF_2}$ crystals exhibit strong anomalous GVD in the mid-infrared regime ($\lambda \sim 2$–20~\textmu m)~\cite{Lin2015,Lecaplain2016}. Excessively strong anomalous GVD requires large Kerr-induced frequency shifts and high pump powers to generate a soliton microcomb. The steady-state solutions of the LLE also predict that the spectral bandwidth decreases in proportion to the square root of the GVD magnitude~\cite{Yi2015,Fujii:20}, as we confirmed in Fig.~\ref{Fig_dispersion_1300}. Since many molecules have specific vibrational modes, referred to as fingerprints, the mid-infrared region is particularly important for sensing applications. Moreover, broadband and spectrally flat combs are feasible for direct frequency comb spectroscopy~\cite{Picque2019}.

\begin{figure*}
\includegraphics[width=0.9\linewidth]{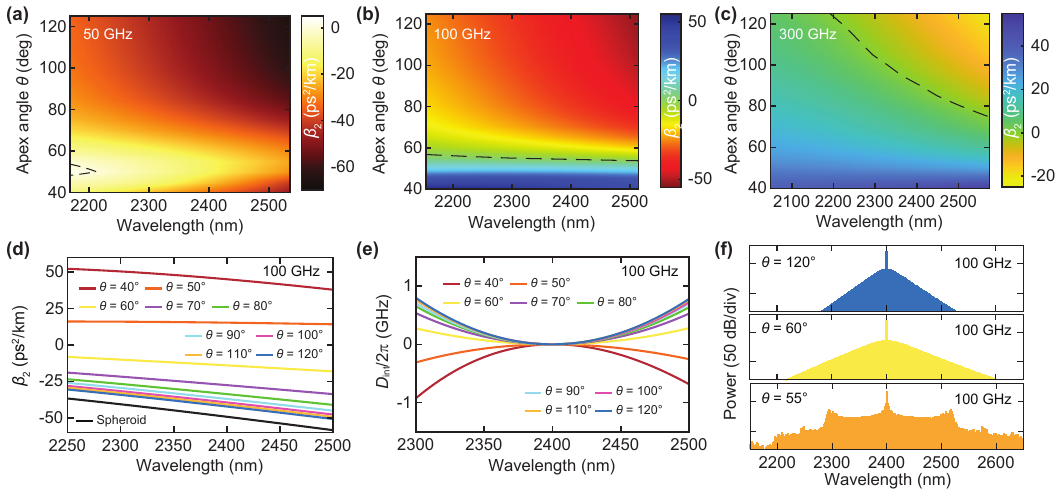}
\caption{\label{Fig_dispersion_2400} (a-c) Colormaps of GVDs with respect to the apex angle $\theta$ for 50, 100, and 300~GHz-FSRs at $\sim$2400~nm. The dashed black lines represent zero dispersion boundaries. (d,e) Simulated GVD $\beta_2$ and corresponding integrated dispersion $D_\mathrm{int}/2\pi$ for 100~GHz FSR resonators. The black line indicates the GVD of the spheroid shape with a curvature $r$=25~\textmu m for comparison. (f) Numerically simulated comb spectra for $\theta=120^\circ$, $\theta=60^\circ$, and $\theta=55^\circ$. }
\end{figure*}

Figure~\ref{Fig_dispersion_1550} again provides an important insight into compensating for strong anomalous GVD by employing a triangular geometry. Figures~\ref{Fig_dispersion_2400}(a)–\ref{Fig_dispersion_2400}(c) show contour maps of the GVD as a function of the apex angle $\theta$ for different FSRs, indicating that optimizing the apex angle enables flexible dispersion control in this wavelength regime. Interestingly, for an FSR of 100~GHz, the strong anomalous GVD ($-40~\mathrm{ps^2/km}$) observed at $\theta = 120^\circ$ reverses its sign near $\theta = 58^\circ$, and eventually becomes strongly normal ($+40~\mathrm{ps^2/km}$) with $\theta = 40^\circ$ at a wavelength of 2400~nm, as shown in Figs.~\ref{Fig_dispersion_2400}(d) and \ref{Fig_dispersion_2400}(e). Simulated microcomb spectra are shown in Fig.~\ref{Fig_dispersion_2400}(f), revealing a distinct transition from bright solitons to a dark-pulse comb as the apex angle decreases. It should be noted that we introduced an additional phase shift into the pump mode to generate a dark-pulse comb in the normal dispersion system~\cite{Xue2015a}. The resulting dark-pulse comb exhibits a flat-top spectrum spanning 200~nm, which is advantageous for spectroscopy compared with $\sim\mathrm{sech}^2$-shaped soliton combs.

\section{Widely tunable optical parametric oscillation pumped at 1~\textmu \lowercase{m} wavelength}

\begin{figure*}
\includegraphics[width=0.9\linewidth]{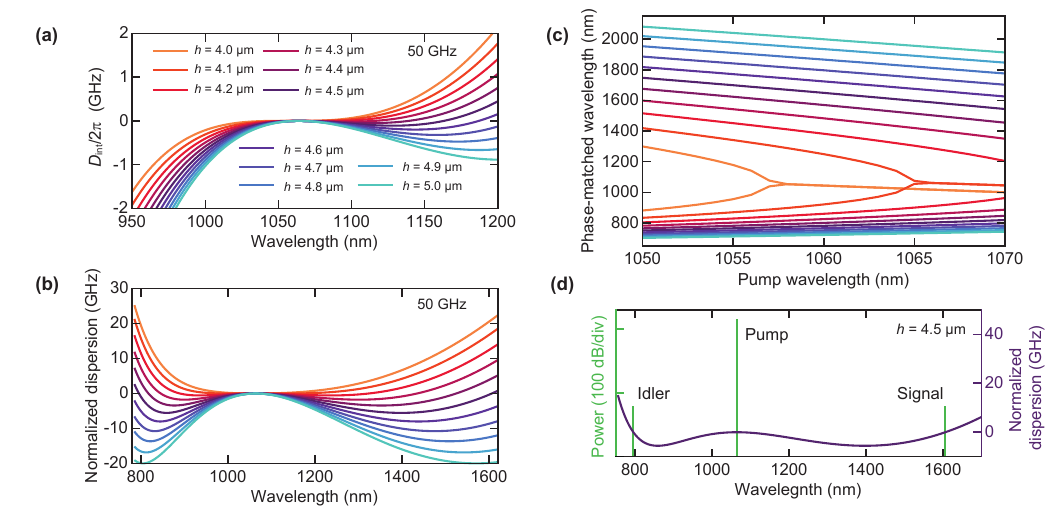}
\caption{\label{Fig_1um_OPO} (a) Integrated dispersion $D_\mathrm{int}$ with an FSR of 50~GHz at a center wavelength of 1064~nm. The colors represent different rectangular heights $h$. (b) Normalized dispersion assumed for the pump wavelength of 1064~nm. The points at which the line crosses zero predict the phase-matched wavelengths. (c) Phase-matched wavelengths as a function of pump wavelength. Degenerate FWM occurs according to the balance between the second and the fourth-order dispersion. (d) Simulated octave-wide OPO spectrum and corresponding normalized dispersion with height $h$=4.5~\textmu m. The signal and idler light are observed at $\sim$800~nm and $\sim$1600~nm when the pump wavelength is 1064~nm.}
\end{figure*}

FWM processes enable not only the generation of soliton frequency combs in anomalous dispersion regimes, but also widely separated OPOs in systems with near-zero normal dispersion~\cite{Sayson2019,Fujii2019}. This unique process has been observed in various photonic resonators and has demonstrated potential for use in quantum photon sources and coherent frequency conversion to the visible and mid-infrared wavelength regimes spanning more than one octave~\cite{Lu2019a,Lu2019}. Because widely separated FWM relies on a delicate balance between second-order and fourth-order dispersion, the phase-matching condition must be carefully controlled. Figures~\ref{Fig_1um_OPO}(a) and \ref{Fig_1um_OPO}(b) show the integrated dispersion for different rectangular waveguide parameters with an FSR of 50~GHz, along with the normalized dispersion, which indicates the approximate phase-matched wavelengths~\cite{Fujii2019}. The normalized dispersion is defined as the frequency difference $\omega_{0+\mu}+\omega_{0-\mu}-2\omega_0 = D_2\mu^2+(D_4/12)\mu^4$. The position of initial parametric sidebands can be approximated from the zero-cross of the value of the frequency difference, yielding,
\begin{equation}
    \mu_{\mathrm{PM}} = \pm \sqrt{\frac{-12D_2}{D_4}},
\end{equation}
where $\mu_{\mathrm{PM}}$ is the mode number of the initial FWM sidebands. Figure~\ref{Fig_1um_OPO}(c) presents the phase-matched wavelength as a function of pump wavelength for waveguide heights $h$ ranging from 4.0~\textmu m to 5.0~\textmu m in 0.1~\textmu m steps. Each color corresponds to the same waveguide geometry shown in Fig.~\ref{Fig_1um_OPO}(a). Widely tunable OPOs spanning 700~nm to 2000~nm are predicted at a pump wavelength of approximately 1064~nm. Despite only slight changes in the structure, the phase-matched wavelengths vary significantly even at the same pump wavelength, as the fourth-order dispersion $D_4$ is highly sensitive to the waveguide geometry, but can nonetheless be engineered. The simulated OPO spectrum is shown in Fig.~\ref{Fig_1um_OPO}(d), where the signal and idler light are generated at octave-separated wavelengths, and bridge the visible to near-infrared spectrum regimes.

\section{Conclusions}

In summary, we have demonstrated systematic dispersion engineering in crystalline $\mathrm{MgF_2}$ WGM microresonators by utilizing ultraprecision machining techniques. This approach enables accurate control over both the GVD and higher-order dispersion parameters, which are critical for tailoring nonlinear optical processes.

By exploring various waveguide geometries, including spheroidal, triangular, rectangular, and trapezoidal cross-sections, we have shown that the dispersion profiles can be flexibly designed to support anomalous or near-zero dispersion regimes across a wide wavelength range. Experimental results confirmed that machine-shaped resonators not only exhibit high geometric precision but also effectively suppress spatial mode interactions, thereby eliminating AMXs that typically hinder soliton generation. Furthermore, we have demonstrated that proper dispersion control allows  widely tunable microcomb generation, ranging from bright solitons in the near-infrared to dark-pulse combs in the mid-infrared regime. We also analyzed phase-matching conditions for broadband OPOs and showed that fourth-order dispersion can be engineered to realize octave-spanning conversion in the 1~\textmu m bands.

Our results highlight the importance of geometric design in achieving low-noise, broadband, and application-oriented frequency combs, and establish ultraprecision machining as a powerful tool for advancing crystalline microresonator-based ultrafast photonics.

\appendix
\section{LLE simulations}

We solved the mean-field LLE, expressed as~\cite{Coen2013}:
\begin{equation}
\begin{split}
\frac{\partial A(\phi,t)}{\partial t} = -\left( \frac{\kappa_{\mathrm{tot}}}{2} + i \delta_0 \right)A -i  \sum_{k=2} \frac{D_k}{k!} \left( \frac{\partial}{i \partial \phi} \right) ^k A \\
+ i g |A|^2 A + \sqrt{\kappa_{\mathrm{ext}}}A_{\mathrm{in}}, \label{Eq.1}
\end{split}
\end{equation}
where $t$ is the time describing the evolution of the field envelope, $\phi $ is the azimuthal angular coordinate. The total decay rate $\kappa_\mathrm{tot}=\omega_0/ Q$ is given by the sum of the intrinsic loss $\kappa_\mathrm{int}$ and the external coupling rate $\kappa_\mathrm{ext}$. $\delta_0=\omega_0-\omega_p$ and $g=\hbar \omega_0^2 c n_2/n V_\mathrm{eff}$ represent the pump detuning from the resonance, and the Kerr nonlinear coefficient, respectively. The nonlinear refractive index $n_2 = 0.9\times10^{-20}$ is used for the simulation. The resonator is driven by a CW input, $A_{\mathrm{in}}=\sqrt{P_{\mathrm{in}}/\hbar \omega_p}$. The dispersion term $D_k$ includes higher-order dispersion, where $k$ represents the index, i.e., $D_2$ is the second-order dispersion. Dispersion and effective mode volume $V_\mathrm{eff}$ are obtained with the FEM simulation~\cite{FujiiTanabe+2020+1087+1104}. We used $\kappa_\mathrm{int}/2\pi=7.1$~MHz, $\kappa_\mathrm{ext}/2\pi=2.3$~MHz, and $P_\mathrm{in}=0.1$~W for Fig.~\ref{Fig_dispersion_1300}(f); $\kappa_\mathrm{int}/2\pi=5.9$~MHz, $\kappa_\mathrm{ext}/2\pi=2.0$~MHz, and $P_\mathrm{in}=0.1$~W for Fig.~\ref{Fig_dispersion_2400}(f). The additional phase term $\Delta$ is added to the pump detuning $\delta_0 \equiv \delta_0+\Delta$ to account for the mode shift, and $\Delta/2\pi=-1$~MHz is used to induce the local anomalous dispersion in the normal dispersion system. For the OPO simulation in Fig.~\ref{Fig_1um_OPO}(d), we used $\kappa_\mathrm{int}/2\pi=1.8$~MHz, $\kappa_\mathrm{ext}/2\pi=1.8$~MHz, and $P_\mathrm{in}=40$~mW.

\section*{Funding}
This work was supported by JSPS KAKENHI (JP24K17624); JST Adaptable and Seamless Technology transfer Program through Target-driven R\&D (A-STEP) (JPMJTR23RF); JST PRESTO (JPMJPR25L9); Keio University Program for the Advancement of Next Generation Research Projects; L. Yang acknowledges China Scholarship Council (202306280149).

\section*{Acknowledgments}
We thank Shinichi Watanabe at Keio Univeristy for the joint use of FEM simulation software.

\section*{Disclosures}
The authors declare no conflicts of interest.

\section*{Data availability} Data underlying the results presented in this paper are obtained from the authors upon reasonable request.


\bibliography{systematic_dispersion}


\end{document}